\documentclass[a4paper
,12pt]{article}
\usepackage{graphicx}
\begin{document}

\title{\bf  A New Statement for Selection and Exception Handling in Imperative Languages}
\author{Keehang Kwon \\
\sl \small Dept. of Computer Engineering, DongA  University\\
\small khkwon@dau.ac.kr
 }
\date{}
\maketitle

\newenvironment{describe}{\begin{list}{}{\setlength\leftmargin{80pt}}\setlength\labelsep{10pt}\setlength\labelwidth{70pt}}{\end{list}}

\newenvironment{flag}{\begin{list}{\makebox[20pt]{\hss$\circ$\enspace}}
                                  {\labelwidth 20pt}}{\end{list}}



\newenvironment{numberedlist}
{\begin{list}{\makebox[20pt]{\hss(\arabic{itemno})\enspace}}
             {\usecounter{itemno}\labelwidth 20pt}}{\end{list}}

\newenvironment{alphabetlist}
{\begin{list}{\makebox[20pt]{\hss(\alph{itemno1})\enspace}}
             {\usecounter{itemno1}\labelwidth 20pt}}{\end{list}}

\newenvironment{romanlist}
{\begin{list}{\makebox[20pt]{\hss(\roman{itemno2})\enspace}}
             {\usecounter{itemno2}\labelwidth 20pt}}{\end{list}}

\newcounter{itemno}

\newcounter{itemno1}

\newcounter{itemno2}
\newcounter{lemma}
\newcounter{exno}

\newcounter{defno}







\newenvironment{defn}{\refstepcounter{defno}\medskip \noindent {\bf
Definition \thedefno.\ }}{\medskip}

\newenvironment{ex}{\refstepcounter{exno}\medskip \noindent {\bf
Example \theexno.\ }}{\medskip}

\newenvironment{millerexample}{
 \begingroup \begin{tabbing} \hspace{2em}\= \hspace{5em}\= \hspace{5em}\=
\hspace{5em}\= \kill}{
 \end{tabbing}\endgroup}

\newenvironment{wideexample}{
 \begingroup \begin{tabbing} \hspace{2em}\= \hspace{10em}\= \hspace{10em}\=
\hspace{10em}\= \kill}{
 \end{tabbing}\endgroup}

\newcommand{\sep}{\;\vert\;}

\newcommand{\ra}{\rightarrow}
\newcommand{\app}{\ }
\newcommand{\appt}{\ }
\newcommand{\tup}[1]{\langle\nobreak#1\nobreak\rangle}

\newcommand{\hu}{{\cal H}^+}
\newcommand{\Free}{{\cal F}}
\newcommand{\oprove}{\vdash\kern-.6em\lower.7ex\hbox{$\scriptstyle O$}\,}
\newcommand{\true}{\top}

\newcommand{\Dscr}{{\cal D}}
\newcommand{\Pscr}{{\cal P}}
\newcommand{\Gscr}{{\cal G}}
\newcommand{\Fscr}{{\cal F}}
\newcommand{\Vscr}{{\cal V}}
\newcommand{\Uscr}{{\cal U}}
\newcommand{\pderivation}{{\cal P}\kern -.1em\hbox{\rm -derivation}}
\newcommand{\pderivationl}{{\cal P}\kern -.1em\hbox{\em -derivation}}
\newcommand{\pderivable}{{\cal P}\kern -.1em\hbox{\rm -derivable}}
\newcommand{\pderivablel}{{\cal P}\kern -.1em\hbox{\em -derivable}}
\newcommand{\pderivations}{{\cal P}\kern -.1em\hbox{\rm -derivations}}
\newcommand{\pderivability}{{\cal P}\kern -.1em\hbox{\rm -derivability}}
\newcommand{\eqm}[1]{=_{\scriptscriptstyle #1}}
\newcommand\subsl{\preceq}
\newcommand{\fnrestr}{\uparrow}

\newcommand{\match}{{\rm MATCH}}
\newcommand{\triv}{{\rm TRIV}}
\newcommand{\imit}{{\rm IMIT}}
\newcommand{\proj}{{\rm PROJ}}
\newcommand{\simpl}{{\rm SIMPL}}
\newcommand{\failed}{{\bf F}}

\newcommand{\Dsiginst}[1]{{[#1]_\Sigma}}
\newcommand{\Psiginst}[1]{{[#1]_\Sigma}}
\newcommand{\lnorm}{{\lambda}norm}
\newcommand{\seq}[2]{#1 \supset #2}
\newcommand{\dseq}[2]{#1_1,\ldots,#1_{#2}}

\newcommand{\all}{\forall}
\newcommand{\some}{\exists}
\newcommand{\lambdax}[1]{\lambda #1\,}
\newcommand{\somex}[1]{\some#1\,}
\newcommand\allx[1]{\all#1\,}

\newcommand{\subs}[3]{[#1/#2]#3}
\newcommand{\rep}[3]{S^{#2}_{#1}{#3}}
\newcommand{\ie}{{\em i.e.}}
\newcommand{\eg}{{\em e.g.}}

\newcommand{\lbotr}{$\bot$-R}
\newcommand{\ldbotr}{\bot\mbox{\rm -R}}
\newcommand{\landl}{$\land$-L}
\newcommand{\ldandl}{\land\mbox{\rm -L}}
\newcommand{\landr}{$\land$-R}
\newcommand{\ldandr}{\land\mbox{\rm -R}}
\newcommand{\lorl}{$\lor$-L}
\newcommand{\ldorl}{\lor\mbox{\rm -L}}
\newcommand{\lorr}{$\lor$-R}
\newcommand{\ldorr}{\lor\mbox{\rm -R}}
\newcommand{\limpl}{$\supset$-L}
\newcommand{\ldimpl}{\supset\mbox{\rm -L}}
\newcommand{\limpr}{$\supset$-R}
\newcommand{\ldimpr}{\supset\mbox{\rm -R}}
\newcommand{\lnegl}{$\neg$-L}
\newcommand{\ldnegl}{\neg\mbox{\rm -L}}
\newcommand{\ldnegr}{\neg\mbox{\rm -R}}
\newcommand{\lalll}{$\forall$-L}
\newcommand{\ldalll}{\forall\mbox{\rm -L}}
\newcommand{\lallr}{$\forall$-R}
\newcommand{\ldallr}{\forall\mbox{\rm -R}}
\newcommand{\lsomel}{$\exists$-L}
\newcommand{\ldsomel}{\exists\mbox{\rm -L}}
\newcommand{\lsomer}{$\exists$-R}
\newcommand{\ldsomer}{\exists\mbox{\rm -R}}
\newcommand{\ldlamlr}{\lambda}
\newcommand{\sequent}[2]{\hbox{{$#1\ \longrightarrow\ #2$}}}
\newcommand{\prog}[2]{\hbox{{$#1\ \supset\ #2$}}}
\newcommand{\run}{\Gamma}

\newcommand{\Ibf}{{\bf I}}
\newcommand{\Cbf}{{\bf C}} 
\newcommand{\Cbfpr}{{\bf C'}}

\newcommand{\cprove}{\vdash_C}
\newcommand{\iprove}{\vdash_I}

\newsavebox{\lpartfig}
\newsavebox{\rpartfig}


\newenvironment{exmple}{
 \begingroup \begin{tabbing} \hspace{2em}\= \hspace{3em}\= \hspace{3em}\=
\hspace{3em}\= \hspace{3em}\= \hspace{3em}\= \kill}{
 \end{tabbing}\endgroup}
\newenvironment{example2}{
 \begingroup \begin{tabbing} \hspace{8em}\= \hspace{2em}\= \hspace{2em}\=
\hspace{10em}\= \hspace{2em}\= \hspace{2em}\= \hspace{2em}\= \kill}{
 \end{tabbing}\endgroup}

\newenvironment{example}{
\begingroup  \begin{tabbing} \hspace{2em}\= \hspace{3em}\= \hspace{3em}\=
\hspace{3em}\= \hspace{3em}\= \hspace{3em}\= \hspace{3em}\= \hspace{3em}\= 
\hspace{3em}\= \hspace{3em}\= \hspace{3em}\= \hspace{3em}\= \kill}{
 \end{tabbing} \endgroup }

\newcommand{\sand}{sand} 
\newcommand{\pand}{pand} 
\newcommand{\cor}{cor} 

\newcommand{\lb}{\langle}
\newcommand{\rb}{\rangle}
\newcommand{\pr}{prov}
\newcommand{\prG}{intp}
\newcommand{\prSG}{intp_E}
\newcommand{\intp}{intp_o}
\newcommand{\prove}{exec} 
\newcommand{\np}{invalid} 
\newcommand{\Ra}{\supset}  
\newcommand{\add}{\oplus} 
\newcommand{\adc}{\&} 
\newcommand{\Cscr}{{\cal C}}
\newcommand{\seqweb}{SProlog}
\newcommand{\sprog}{{SProlog}}

\newtheorem{theorem}[lemma]{Theorem}

\newtheorem{proposition}[lemma]{Proposition}

\newtheorem{corollary}[lemma]{Corollary}
\newenvironment{proof}
     {\begin{trivlist}\item[]{\it Proof. }}%
     {\\* \hspace*{\fill} \end{trivlist}}

\newcommand{\seqand}{\prec}
\newcommand{\seqor}{\cup}
\newcommand{\seqandq}[2]{\prec_{#1}^{#2}}
\newcommand{\parandq}[2]{\land_{#1}^{#2}}
\newcommand{\exq}[2]{\exists_{#1}^{#2}}
\newcommand{\ext}{intp_G}

\newcommand{\muprolog}{{Java$^{ch}$}}
\newcommand{\kch}{seqor}
\newcommand{\kcha}{seqand}
\renewcommand{\pr}{ex}
\renewcommand{\prove}{ex} 
\newcommand{\iif}{$if$-$then$-$else$}
\newcommand{\swi}{$switch$}
\newcommand{\try}{$try$-$catch$}

\noindent {\bf Abstract}: 
Diverse selection statements -- \iif, \swi\ and
\try -- 
are commonly used  in modern programming languages. 
To make things simple, we propose a unifying  statement for selection. This statement 
is of the form  $\kch(G_1,\ldots,G_n)$ where each  $G_i$ is a statement.  It has 
 a simple  semantics:   sequentially choose the first $successful$ statement
$G_i$ and then proceeds with executing $G_i$.                    
Examples will be provided for this  statement.

{\bf keywords:} selection, imperative programming, exceptions

\section{Introduction}

Most programming languages have selection statements to direct execution flow.
A selection statement allows the machine to 
 choose one between two or more tasks during execution.
Selection statements include \iif, \swi, \try\ and their variations.
 Unfortunately, these statements were designed on an ad-hoc basis
 and have several shortcomings.
 
One big design flaw with imperative languages seems that 
 $true/false$ and $success/failure$ are treated differently. This causes a lot of
complications in programming style. In addition, this leads to two separate control
       statements: one for control statement and one for exception handling statement.

Our approach to overcoming these problems is the following:
 a boolean condition is considered a legal statement
in our language. For example, $2==3$, $prime(6)$  return failure rather than false.
In this setting, $false$ is replaced by $failure$, leading
 to a single selection statement  for both control and exception handling.

  We now interpret each statement as T/F(success/failure),
depending on whether it has been successfully completed or not.
Note that our semantics is based  on a $task$-$logical$ approach (see, for example, \cite{Jap03,Jap08}.
Our work is in fact motivated by sequential operators in \cite{Jap08}.)
 to exception handling because it includes the notion of success/failure. 
In this setting, every exception(including false) is interpreted as failure.

In this setting, we propose a  new  selection  statement.
This statement  is  of the form

\[ \kch(G_1,\ldots,G_n) \]
\noindent
where each  $G_i$ is a statement. This has the following execution semantics:

\[ \pr(\Pscr, \kch(G_1,\ldots,G_n))\ if\
 \pr(\Pscr,  G_i) \] 

\noindent where  $\Pscr$ is a 
set of procedure  definitions.
In the above definition,  the machine sequentially chooses the first  $successful$ disjunct
$G_i$ and then proceeds
with executing $G_i$.  That is, the
machine {\em sequentially} tries these statements  from left to
right. 

Our  $\kch$ statement has several  uses:

\begin{numberedlist}

\item Suppose this statement is of the form  \\
  $\kch(cond_1;G_1,\ldots,cond_n;G_n,G_{n+1})$. Suppose also that
 each $G_i$ is guaranteed to terminate successfully.
 It can be easily seen that our  statement is a  simpler alternative to
the old \iif\ statement in traditional imperative languages such as  C. 

For example, $if\ cond\ then\ S\
else\ T$ can be converted to

\[ \kch(cond; S, T). \]
\noindent and vice versa.

It is also straightfoward to convert the \swi\ statement
to our language. For example,
the following Java-like code displays  the employee's age. 

\begin{exmple}
        getAge(emp) \{ \\
        switch (emp) \{ \\
 \>           case tom:  age = 31;   break; \\
  \>          case kim: age = 40;   break;\\
 \>           case sue: age = 22;    break;\\
 \>           default: age = 0;      \\
        \}\\
 return age; \}
\end{exmple}

\noindent 
 Note that the above code  can be converted to the one
below:
 
\begin{exmple}
       getAge(emp) \{ \\
        \kch( \\
 \>           emp == tom;  age = 31, \\
  \>          emp == kim; age = 40, \\
 \>           emp == sue; age = 22, \\
\>            age = 0 ); \\
return age; \} 
\end{exmple}

\noindent This program expresses the task of the machine sequentially choosing one among
three employees.
 Note that this program is  compact and
 easier to read. 

\item Suppose, in the above, some $G_i$ terminates unsuccessfully. 
Then the above statement executes the next statement $cond_{i+1};G_{i+1}$.
Thus it behaves differently from the old \iif . This semantics is natural, as we shall
see later in the DFS tree example.

\item Suppose this statement is of a general form $\kch(G_1,\ldots,G_n)$ where
each $G_i$ is an arbitrary  statement. 
It can then be observed that this statement is well-suited to
exception handling. That is, $G_2$ is intended to handle exceptions raised in $G_1$,
$G_3$  to handle exceptions raised in $G_2$, and so on.

This statement is a
 simpler alternative to the traditional \try\ statement. The main differences
between these two are  the following:

\begin{itemize}

\item Most importantly, our statement has a simpler syntax and semantics. For example, our statement associates
exceptions and handlers statically.

\item  While the \try\ statement is designed as a binary connective, the $\kch$ statement
    is designed as an 
$n$-ary connective.

\item  Now every exception  belongs to a single parent called F. That is, F is the parent of
all exceptions. This hierarchical structure makes exception handling
     simpler.


\item Our language supports the strict separation of (a)~ error recovery and restart to be handled by
the machine and (b)~ alternative tasks to be specified by the programmer. To be specific,
if $G_j$ fails in the middle of executing the $\kch(G_1,\ldots,G_n)$ statement,  
 then it is required that
the machine -- not the programmer --  performs the recovery and restart action. In other
words, it rolls back partial updates caused by $G_j$. This is typically done in
$G_{j+1}$  in traditional exception handling mechanisms.
As a result,  $G_{j+1}$ becomes simpler in our language,
 because it only needs to specify  alternative tasks to do.

\end{itemize}

\end{numberedlist}

This paper focuses on the minimum 
core of Java. This is to present the idea as concisely as possible.
The remainder of this paper is structured as follows. We describe 
our language
 in Section 2. In Section \ref{sec:modules}, we
present an example of  \muprolog\ that deals with exception handling.
Section~\ref{sec:conc} concludes the paper.

\section{The Language}\label{sec:logic}

The language is a subset of the core (untyped) Java
 with some extensions. It is described
by $G$- and $D$-formulas given by the syntax rules below:
\begin{exmple}
\>$G ::=$ \>   $t \sep f \sep A \sep cond\sep \neg cond
\sep x = E \sep  G;G \sep   \kch(G_1,\ldots,G_n)$ \\   \\
\>$D ::=$ \>  $ A = G\ \sep \all x\ D$\\
\end{exmple}
\noindent
In the above, $t$ represents a (user-defined) success and $f$ represents
a (user-defined) failure/exception. $f$ is often extended to
$f(errcode)$ which is used to raise an exception. In addition, $x$ represents a variable and
$A$  represents a head of an atomic procedure definition of the form $p(x_1,\ldots,x_n)$.
A $D$-formula  is called a  procedure definition.

In the transition system to be considered, $G$-formulas will function as the
main program (or statements), and a set of $D$-formulas enhanced with the
machine state (a set of variable-value bindings) will constitute  a program.
Thus, a program consists of a union of two disjoint sets, \ie, $\{ D_1,\ldots,D_n \} \cup \theta$
where each $D_i$ is a $D$-formula and $\theta$ represents the current machine state.
 $\theta$ is initially an empty set and will be updated dynamically 
via the assignment statements. 

 We will  present an operational
semantics for this language via a proof theory \cite{Khan87}. This style of
semantics has been used in logic languages \cite{MNPS91,HM94,MN12}.
Note that the machine  alternates between two phases:
 the execution phase 
and the backchaining phase.  
In  the execution phase (denoted by $ex(\Pscr,G,\Pscr')$) it  executes a main statement $G$ relative to
a program $\Pscr$ and
produces a new program $\Pscr'$
by decomposing $G$ -- via  rules
  (8) and (9) --
to simpler forms until $G$ becomes an assignment statement, a conditional statement,
 or a procedure call. 
If $G$ becomes a procedure call, the interpreter switches to the backchaining mode (via rule (3)).
In the backchaining mode (denoted by $bc(D,\Pscr,A,\Pscr')$), the interpreter tries 
to solve a procedure call  $A$ and produce a new  program $\Pscr'$
by first reducing a procedure definition $D$ in a program $\Pscr$ to  its instance
 (via rule (2)) and then backchaining on the resulting 
definition (via rule (1)).
 The notation $S$\ seqand\ $R$ denotes the  sequential execution of two tasks. To be precise, it denotes
the following: execute $S$ and execute
$R$ sequentially. It is considered a success if both executions succeed.
Similarly, the notation $S$\ parand\ $R$ denotes the  parallel execution of two tasks. To be precise, it denotes
the following: execute $S$ and execute
$R$  in any order.  
It is considered a success if both executions succeed.
The notation $S \leftarrow R$ denotes  reverse implication, \ie, $R \rightarrow S$.

\begin{defn}\label{def:semantics}
Let $G$ be a main statement and let $\Pscr$ be a program.
Then the notion of   executing $\lb \Pscr,G\rb$ successfully and producing a new
program $\Pscr'$-- $ex(\Pscr,G,\Pscr')$ --
 is defined as follows:

\begin{numberedlist}

\item    $bc((A = G_1),\Pscr,A,\Pscr_1)\ \leftarrow$ 
 $ex(\Pscr, G_1,\Pscr_1)$. \% A matching procedure for $A$ is found.

\item    $bc(\all x D,\Pscr,A,\Pscr_1)\ \leftarrow$    $bc([s/x]D,
\Pscr, A,\Pscr_1)$ where $s$ is a term. \% argument passing

\item    $ex(\Pscr,A,\Pscr_1)\ \leftarrow$    $(D \in \Pscr$ parand $bc(D,\Pscr, A,\Pscr_1))$. \% a procedure call

\item  $ex(\Pscr,t,\Pscr)$. \% True is always a success.

\item  $ex(\Pscr,cond,\Pscr)$ if $cond$ is true. \% boolean condition.

\item  $ex(\Pscr,\neg cond,\Pscr)$ if $cond$ is false. \% boolean condition
.
\item  $ex(\Pscr,x=E,\Pscr\uplus \{ \lb x,E' \rb \})\ \leftarrow$  $eval(\Pscr,E,E')$.
\% the assignment statement. Here, 
$\uplus$ denotes a set union but $\lb x,V\rb$ in $\Pscr$ will be replaced by $\lb x,E' \rb$.

\item  $ex(\Pscr,G_1; G_2,\Pscr_2)\ \leftarrow$   $(ex(\Pscr,G_1,\Pscr_1)$  seqand 
  $ex(\Pscr_1,G_2,\Pscr_2))$. \% sequential composition

\item $ex(\Pscr, \kch(G_1,\ldots,G_n), \Pscr_1)$  if 
$ex(\Pscr, G_i,\Pscr_1)$ where $G_i$ is the first successful 
statement.

\end{numberedlist}
\end{defn}

\noindent
If $ex(\Pscr,G,\Pscr_1)$ has no derivation, then the machine returns  the failure. For example, $ex(\Pscr,F,\Pscr_1)$ is a failure
because there is no derivation for $f$.

 The rule (9) deals with selection.    To execute $\kch(G_1,\ldots,G_n)$ successfully, 
the machine does the following:

\begin{numberedlist}

\item The machine tries $G_1$. If it returns a success, then the executin terminates
with a success. If it returns the failure, then it performs the recovery 
action by rolling back partial updates caused by $G_1$, and then tries $G_2$. It goes on
until the machine finds some $G_j$ that leads to a success.

\item If all $G_i$s fail, the machine returns the failure with a list of error codes.

\end{numberedlist}
\noindent
As mentioned earlier, the $\kch$  construct is a well-designed, 
high-level abstraction for selection and exception handling.

\section{Examples }\label{sec:modules}

The traditional imperative approach is inadequate for representing failure.
For example, consider a DFS search algorithm. There is no  uniform way to represent
and handle the case when the search terminates unsuccessfully.
Different codes returns different values such as 0, false, nil, exception, etc.
As a consequence, the resulting code becomes very awkward.

We now describe a procedure $dfs(tree,key)$ which searches for a $key$ in a binary tree $T$.
$nil$ represents an empty tree.

\begin{exmple}
 procedure       $ dfs((r,L,R),key)$ \{ \% root,left subtree, right subtree \\
         \kch\ ( \\
 \>           $r == key$,   \% found \\
  \>          $L \neq nil; dfs(L,key)$,  \% search left subtree \\
 \>          $R \neq nil; dfs(R,key)$  \% search right subtree \\
        )\\
\end{exmple}
\noindent 
In case the key is not found, the above code simply fails.
Note that the above code is as concise as possible, requiring no special code for unsuccessful
termination of search.

As another example of exception handling, let us consider the following three tasks.

A: Send a message $m$ via $send\_fast(m)$ which is normally the better way to send a
 message, but it may fail, triggering an exception.

B: Send a message $m$ via $send\_slow(m)$ which  will fail less often.

C: Send a message $m$ via $send\_slowest(m)$ which  will hardly fail.

\noindent It is not easy to write robust codes for these tasks in
traditional languages.
Fortunately, it is rather simple in our setting. For example,
the following statement expresses the task of trying A, B and C sequentially.

\[ \kch(A,B,C) \]

\noindent In the above, if all  fail, the machine returns the failure with a list
of error codes.
Of course, a further analysis of exceptions is possible by inspecting
the exceptions raised during execution. 
In summary, our language considerably reduces many complications.

\section{Negative Exception Handling}\label{sec:neh}

It can be easily observed that  exception handling in the main program
has a dual notion. That is, exception handling
is possible at the declaration program via seq-and procedures. For example, the overloaded
procedure $draw$ can be written as follows:
\begin{exmple}
  \kcha\ ( \\
  $      draw(X:circle)\ =$  \> \hspace{8em}   $\ldots$, \% draw a circle \\
  $      draw(X:point)\ =$  \> \hspace{8em} $\ldots$,   \% draw a point \\
  $      draw(X:rectangle)\ =$  \> \hspace{8em} $\ldots$  \% draw a rectangle\\
  ) 
\end{exmple}
\noindent In this setting, the machine tries the first procedure . If it fails, then it tries the next
and so on.
This aspect -- which we call {\it negative} exception handling --
was discussed  in the context of functional languages\cite{Kfunc}.

\section{Conclusion}\label{sec:conc}

In this paper, we have considered an extension to a core Java with a new selection statement.
 This extension allows   $\kch(G_1,\ldots,G_n)$  where each $G_i$ is a statement.
This statement makes it possible for the core Java
to deal with exceptions as simply as possible.  

Exception handling is quite  challenging because  softwares
 are getting  more and more complex. 
Unfortunately, exception handling in modern programming languages 
has quite unsatisfactory: 
 The design and analysis of exception code is quite complicated.
Therefore, the need for a new exception handling mechanism is clear.

Our $\kch$ statement is well-suited to  exception handling.

\begin{itemize}

\item It has a simple syntax and semantics.

\item Our language naturally gives a $hierarchical$  structure to exceptions.
       Now all the exceptions belong to the Failure set. This 
       considerably simplifies the exception handling.
   
\item Our language disallows abrupt, nonlocal, dynamic
 transfers of control which
          violates the declarative reading of the execution sequence.
          This property is essential for program verification.
        
 \item Our language can easily handle $nested$ exception handling. 
 
 \item Our language gives a logical status to exceptions. This means that other useful logical connectives such as disjunctions can be
 added. Some progress has been made towards this direction \cite{KHP13}.

 \end{itemize}



\bibliographystyle{ieicetr}



\end{document}